\documentclass[aps,pra,superscriptaddress,twocolumn]{revtex4}
\usepackage{amsmath,epsfig,mathrsfs,graphicx,amssymb,color}
\usepackage[T1]{fontenc}
\usepackage{t1enc}
\usepackage{hyperref}
\def\be#1\ee{\begin{equation}#1\end{equation}}
\newcommand{\ba}{\begin{eqnarray} }
\newcommand{\ea}{\end{eqnarray} }

\DeclareMathOperator\sinc{sinc}

\begin{document}

\title{Nonclassical correlations in decaying systems}
\author{Stanis{\l}aw So{\l}tan}
\author{Adam Bednorz}
\affiliation{Faculty of Physics, University of Warsaw, ul. Pasteura 5, PL02-093 Warsaw, Poland}
\email{Adam.Bednorz@fuw.edu.pl}

\date{\today}

\begin{abstract}
A quantum decaying system can reveal its nonclassical behavior by being noninvasively measured. 
Correlations of weak measurements in the noninvasive limit violate the classical bound for a universal class of systems.
The violation is related to the incompatibility between exponential decay and unitary evolution.
The phenomenon can be experimentally observed by continuous weak measurements and a large class of observables.
The nonclassical nature of such a system allows us to treat it as a potential quantum resource.

\end{abstract}

\maketitle

\section{Introduction}

Quantum objects from atoms, nuclei, and condensed matter systems to high energy particles often exhibit exponential decay despite the underlying unitary dynamics, commonly derived by Fermi golden rule \cite{dirac,fermi}.
It follows from the generic structure of the decaying system, a local state coupled to an open reservoir.
The decay can be slowed or accelerated by frequent or continuous measurements \cite{kofman,elattari,koshino2004,koshino2005,giaco}. A very frequent or strong measurement
can lead to complete freezing of the decay, known as the quantum Zeno effect \cite{zeno}. Nevertheless, except for some extreme cases \cite{winter}, the decay is generally exponential in time.

The Zeno effect makes it clear that one cannot simply monitor the decay, especially by  means of invasive measurements.
The projective measurement is always invasive even if it is rarely performed. In this case, the decay rate may remain almost unaffected
but projection destroys the fragile coherence. Unfortunately, the effect of the measurement on the decay rate gives no quantitative conclusion on the nonclassicality of
the decay. The measurement can either slow down or speed up the decay while the system should exhibit quantum properties in both cases. Moreover, the correlations between subsequent projective measurements reveal just the initial
formation of the decaying state wavefunction. To prove the quantum nature of such systems, the measurement cannot destroy coherence. 
A natural solution is a weak measurement, i.e. a weakly coupled detector \cite{aav}, realized already also in decaying systems \cite{weakd1,weakd2}.
Weak measurements are the strongest candidates for noninvasive measurements, as the disturbance is as small as the square of the detector-system coupling,
preserving the coherence of the system.
This applies to both closed and open systems.
On the other hand, a weak measurement suffers from a large noise, inevitable in any attempt to extract information about the evolution of the quantum system noninvasively. 

Here we show that the correlations
of weak measurement can reveal nonclassical features of quantum decaying systems in analogy to anomalous weak values or violations of various
classical Leggett-Garg-type inequalities \cite{lega,emary,palacios-laloy:10,matz,nori0,nori,lld,lgspin1,lgspin2,energy}. The correlations exceed
the classical limit, if results of noninvasive measurement are treated as an objective element of reality,  
removing the inevitable detection noise \cite{bbn,bfb}. 
Whether the removal of this noise is a correct step to identify the quantum objective reality, 
is a matter of debate \cite{leg,fercom,com,sokol,vaidman}. An equally admissible interpretation is to include all the detection noise, making all classical inequalities valid. However, classical systems could also be measured noninvasively, with a large noise added, so its removal seems to be a fair assumption to discriminate between classical and quantum systems. 
To our knowledge, Leggett-Garg-type inequalities yet to be investigated for decaying systems.
The violation is a universal
feature for sufficiently short times between measurements, given that the measured observable is a projection, at least approximately.
It follows from the incompatibility of the exponential decay and unitary dynamics between the measurements.
Our observation is complementary to the freezing of the dynamics by   spoiling the coherence due to strong measurements
 but is more universal, occurring in principle in every decaying system.
It confirms quantitatively that the quantum decay cannot be explained by a simple classical relaxation process and the decaying system is 
a genuine quantum information resource.

We illustrate our findings on specific decay models. We propose the natural implementation by a continuous weak monitoring of the state.
The effect of the finite measurement strength can be estimated or derived analytically for particular simple models. Our analysis involves a perturbative
expansion in the coupling between the state and the reservoir. Some predictions are universal to be applied to specific experimental
setups.

\section{Weak measurement of a decaying state}

The decaying system transfers the amplitude of the remaining local state into the continuous spectrum of reservoir states. 
The local state is fully occupied initially and empty finally. At times longer than the initial timescales, its amplitude decays essentially exponentially.
Such a pseudostationary state $|\psi\rangle$ can be effectively described as an eigenstate of a Hermitian (self-adjoint in $L^2$ space) Hamiltonian $\hat{H}$ but (i) with a complex eigenvalue  $-i\Gamma$ and (ii) non-normalizable wavefunction (diverging in the reservoir), 
\begin{equation}
\hat{H}|\psi(t)\rangle=i\partial_t|\psi(t)\rangle=-i\Gamma|\psi(t)\rangle,\label{gam}
\end{equation}
 using the convention $\hbar=1$. Later, we will find $\Gamma$ exactly for a particular model of $\hat{H}$. 
This asymptotic description is correct as we can ignore the parts of the reservoir that are far away from the local state so that 
they can no longer interact with this state.
Therefore, our generic starting point is the equation. For a specific $\hat{H}$, there exists a method to replace the problem by a few-level system and non-Hermitian Hamiltonian, using pseudomodes \cite{pseudo}. Nevertheless, (\ref{gam}) is true for a  general class of Hermitian (self-adjoint) Hamiltonians and the following  general discussion is not restricted to any specific case, except later examples.
For a moment it suffices to assume that $R=2\mathrm{Re}\Gamma>0$, i.e. there is indeed some exponential decay.
The time dependence reads $|\psi(t)\rangle=e^{-\Gamma t}|\psi\rangle$. One should not be worried about the lack of normalization, we always refer to some
$t=t_0\gg 1/\Gamma$ and the probability below is simply scaled by $e^{-\Gamma t_0}$.

The sequence of measurement of the same observable $\hat{A}$  at times $t_1<t_2<...<t_N$ with outcomes $a_1,a_2,...,a_N$ can be described by a sequence of 
Kraus operators \cite{kraus}
\begin{equation}
\hat{K}(a)=\hat{K}_N(a_N,t_N)\cdots\hat{K}_2(a_2,t_2)\hat{K}(a_1,t_1)
\end{equation}
giving the probability $P(a_1,a_2,...,a_N)=\mathrm{Tr}\hat{K}(a)\hat\rho\hat{K}^\dag(a)$ for the state $\hat{\rho}$ ($=|\psi\rangle\langle\psi|$ 
in the case of a pure state $|\psi\rangle$). We choose the natural Gaussian form  $\hat{K}=(\pi/4g)^{-1/4}\exp[-2g(a-\hat{A}(t))^2]$
with  Heisenberg observable $\hat{A}$.
In the continuous limit $g\to \lambda\Delta t$ and $t_j\to j\Delta t$ we get \cite{hel,peres,kraus,wiseman},
\begin{equation}
\hat{K}[a]=Z\mathcal T\exp\int dt \left[-2\lambda(a(t)-\hat{A}(t))^2\right]
\end{equation}
with the measured time trace (continuous limit of a sequence of outcomes) $a_1,a_2,...a_N\to a(t)$, denoting time ordering (later to the left of the earlier one) in the expansion of the exponential $\mathcal T$ ,
$Z$ -- the normalization factor, $\lambda>0$ -- the strength of the measurement with $\lambda \to \infty$ being the strong, freezing measurement,
while $\lambda \to 0$ being the continuous weak, noninvasive measurement.

The probability becomes a functional, $P(a_1,a_2,...,a_N)\to P[a]$, which
is a convolution
\begin{equation}
P[a]=\int D\bar{a} N(a-\bar{a})Q(\bar{a})
\end{equation}
of a pure white detection noise $N$, Gaussian, zero-centered, with correlations $\langle a(t)a(t')\rangle_N=\delta(t-t')/8\lambda$  and the special quantum measure $Q$, which turns out to be a quasiprobability (sometimes negative), giving for $\lambda\to 0$ the correlations in the form of nested anticommutator $\check{A}\hat{X}=\{\hat{A},\hat{X}\}/2$ \cite{bfb},
\begin{equation}
\langle a(t'')\cdots a(t')a(t)\rangle_Q=\mathrm{Tr}\check{A}(t'')\cdots\check{A}(t')\check{A}(t)\hat{\rho}\label{abz}
\end{equation}
for the sequence of times $t''>...t'>t$. 
To keep the detection noise contribution finite, the measurement must be regularized, i.e. $\tilde{a}(t)=\int a(t')g(t-t')dt'$ 
with some spread $g$, giving $\langle \tilde{a}^2\rangle_N=\int g^2(t)dt/8\lambda$.

\section{Violation of Leggett-Garg-type inequality}

Now let $\hat{A}$ be a local projection on the unstable state, i.e., $\hat{A}^2=\hat{A}$, and it is nonzero only locally around the position of the decaying state. Since the objective value of $\hat{A}\to a$ is limited to $[0,1]$, we have the classical bound on correlations
\begin{equation}
\langle a(t)a(0)\rangle\leq \langle a(t)\rangle.\label{clas}
\end{equation}
The quantum average, in the limit $\lambda\to 0$ reads
\begin{equation}
\langle a(t)\rangle_Q=\langle A(t)\rangle=e^{-Rt}\langle \psi|\hat{A}|\psi\rangle
\end{equation}
On the other hand the correlation of weak measurements for $\lambda\to 0$ reads
\begin{equation}
\langle a(t)a(0)\rangle_Q=\langle \psi|\{\hat{A}(t),\hat{A}(0)\}|\psi\rangle/2,
\end{equation} 
with the anticommutator $\{\hat{X},\hat{Y}\}=\hat{X}\hat{Y}+\hat{Y}\hat{X}$.
It is calculated by
expanding $A(t)=e^{i\hat{H}t}\hat{A}e^{\hat{H}t/i}$,
\begin{eqnarray}
&&\langle a(t)a(0)\rangle=\mathrm{Re}\;e^{-\Gamma^\ast t}\langle\psi|\hat{A}e^{\hat{H}t/i}\hat{A}|\psi\rangle\nonumber\\
&&= e^{-Rt/2}(\langle a(0)\rangle+O(t^2))
\end{eqnarray}
i.e. the term linear in $t$ vanishes for small $t$.
Therefore for sufficiently low $t$ we obtain violation of the classical bound (\ref{clas}) on the conditional average
\begin{equation}
\frac{\langle a(t)a(0)\rangle_Q}{\langle a(t)\rangle_Q}\equiv \langle a(0)||a(t)\rangle_Q\simeq e^{Rt/2}>1\label{cor}
\end{equation}
where $\langle \cdot||\cdot\rangle$ denotes the conditional average defined on the left.
This is our main result, the violation of \ref{clas} when replacing standard probability by $Q$, which is a Leggett-Garg (LG)-type inequality \cite{lega}. 
Mathematically, the violation is caused by the correlation  $\sim e^{-Rt/2}$ while
the average is $\sim e^{-Rt}$ since the projection pushes the state back to $a=1$ and it starts from the unitary dynamics, in analogy
to the Zeno effect for frequently repeated projections. Note, however, that here the measurement is weak and the decay is not slowed down or slowed minimally
in contrast to the traditional Zeno effect.

For a more general observable, using twice the Cauchy-Bunyakovsky-Schwarz inequality $\langle XY\rangle^2\leq \langle X^2\rangle\langle Y^2\rangle$
we can construct another LG-type inequality
\begin{equation}
\langle a^3(t)a(0)\rangle^4\leq \langle a^4(t)\rangle^3\langle a^4(0)\rangle.\label{acor}
\end{equation}
On the other hand, a quantum calculation for the decaying state gives in the noninvasive limit
\begin{equation}
\langle a^4(t)\rangle_Q=\langle \psi|\hat{A}^4(t)|\psi\rangle=e^{-Rt}\langle \psi|\hat{A}^4|\psi\rangle\label{ah1}
\end{equation}
while the correlation of weak measurements reads
\begin{eqnarray}
&&\langle a^3(t)a(0)\rangle_Q=\langle \psi|\{\hat{A}^3(t),\hat{A}(0)\}|\psi\rangle/2=\nonumber\\
&&
\mathrm{Re}\; e^{-\Gamma^\ast t}\langle \psi|\hat{A}^3e^{\hat{H}t/i}\hat{A}|\psi\rangle\simeq\label{aha}\\
&&
e^{-Rt/2}(\langle \psi|\hat{A}^4|\psi\rangle+t\langle\psi|[\hat{A}^2,\hat{A}\hat{H}\hat{A}]|\psi\rangle/2i),\nonumber
\end{eqnarray}
neglecting $t^2$ terms in the last parentheses (we used the commutator $[\hat{X},\hat{Y}]=\hat{X}\hat{Y}-\hat{Y}\hat{X}$).
For projections the second term vanishes and the inequality (\ref{acor}) is violated as $e^{-2Rt}\geq e^{-3Rt}$. 
Other cases of vanishing of the second term include specific forms of $\hat{H}$ and $\hat{A}$, e.g. $\hat{H}$ is diagonal is the basis of the nonzero
part of $\hat{A}$.

\section{Measurement of finite strength}

For a finite $\lambda$,
in the closed system, a continuous measurement makes decoherence, leading ultimately to heating, so the state is weakly affected only at sufficiently
short times compared to the timescale $\lambda^{-1}$. In the case of open systems, including the decaying ones, they either reach some stationary mixed state
or the rate of decay is modified \cite{weakd2}. To estimate the change of the decay in weak measurements, 
let us ignore the outcome $a(t)$ and check the effect of the measurement on the evolution. The quantum state $\hat{\rho}(t)$ 
undergoes then the Lindblad equation \cite{kos,lin,bbn}
\begin{equation}
\partial_t\hat{\rho}=\check{L}\hat{\rho}=[\hat{H},\hat{\rho}]/i-\lambda[\hat{A},[\hat{A},\hat{\rho}]].\label{lind}
\end{equation}
Without the measurement, i.e. $\lambda=0$ we have $\hat{\rho}(t)=e^{-Rt}|\psi\rangle\psi|$. For any finite measurement strength, we have
$\hat{\rho}(t)=e^{-R't}\hat{\rho}'$. The derivation of the corrections $R\to R'$ and $|\psi\rangle\langle \psi|\to\hat{\rho}'$ is a bit tricky because
the states $|\psi\rangle$ and $\hat{\rho}'$ are not normalized -- they diverge in the reservoir space. To adjust the standard perturbative approach and 
avoid infinities, we need a technical trick. We introduce the antidecaying/absorbing state $|\tilde{\psi}\rangle$ \cite{hart} such that 
\begin{equation}
\hat{H}|\tilde{\psi}\rangle=+i\Gamma^\ast|\tilde{\psi}\rangle.\label{ant}
\end{equation}
The existence of this state follows from the Hermiticity of $\hat{H}$ and simple intuition -- instead of the decay, we reverse the process, letting the 
central state accumulate, absorbing the amplitude from the reservoir. Just as the decaying state, it is divergent in the reservoir but in the opposite
part, so that the overlap $\langle \tilde{\psi}|\psi\rangle$ is finite. We shall see it later in detail on a simple example. Now, we can construct the formal
perturbative expansion in $\lambda$. Sandwiching (\ref{lind}) with $\tilde{\psi}$, we get
\begin{equation}
(R'-R)\langle \tilde{\psi}|\hat{\rho}'|\tilde{\psi}\rangle=\lambda\langle\tilde{\psi}|[\hat{A},[\hat{A},\hat{\rho}']]|\tilde{\psi}\rangle/2
\end{equation}
giving in the lowest order
\begin{equation}
R'-R=2\lambda\left(\mathrm{Re}\frac{\langle\tilde{\psi}|\hat{A}^2|\psi\rangle}{\langle\tilde{\psi}|\psi\rangle}-
\frac{|\langle\tilde{\psi}|\hat{A}|\psi\rangle|^2}{|\langle\tilde\psi|\psi\rangle|^2}\right)
\end{equation}
It confirms that the weak measurement does not change the decay rate considerably. As regards higher orders, one has to solve the Lindblad equation
(\ref{lind}) perturbatively with appropriate boundary conditions, diverging in the reservoir). Due to this divergence, one cannot make the decomposition into
unitary eigenstates. It is somewhat analogous to the Stark effect, where it is more convenient to solve the differential equation directly than
sum over hydrogen bound and scattering states \cite{landau}.

To include the finite $\lambda$ effect, one has to insert the evolution (\ref{lind}) into the formulas for correlations of weak measurement \cite{bbn}, i.e.
\begin{eqnarray}
&&\langle a^n(t)\rangle_Q=\mathrm{Tr}\hat{A}^n e^{t\check{L}}\hat{\rho},\nonumber\\
&&\langle a^n(t)a(0)\rangle_Q=\mathrm{Tr}\hat{A}^n e^{t\check{L}}\check{A}\hat{\rho}
\end{eqnarray}
if $t\geq 0$ (compare with (\ref{abz})). Therefore, the violation of (\ref{clas}) and (\ref{acor})  can be checked also for finite $\lambda$.

\section{Examples}

Let us now illustrate our results on a simple example. The space consists of the single state $|\Omega\rangle$, measured by $\hat{A}=|\Omega\rangle\langle \Omega|$ which decays into the continuum of states
$|x\rangle$, $x\in\mathbb R$, with normalization $\langle x|y\rangle=\delta(x-y)$. In principle we could switch into momentum space 
$|p\rangle=\int e^{ipx}|x\rangle$ by the divergence of the decaying state making such a transform ill-defined.
The Hamiltonian reads
\begin{eqnarray}
&&\hat{H}=\int dx \bar{V}(x)|x\rangle\langle \Omega|+\mathrm{h.c.}+\hat{p},\nonumber\\
&&\hat{p}=\int \delta'(x-y)|x\rangle\langle y|dxdy/i\label{hamp}
\end{eqnarray}
with some complex interaction potential $\bar{V}(x)$.  The momentum part $\hat{p}$ simply moves the state in the positive $x$ direction at velocity $1$
(by our convention both $x$ and $t$ are dimensionless), i.e. $|x\rangle\to |x+t\rangle$. The interaction transfer the amplitude from $|\Omega\rangle$ to $|x\rangle$ and moves away towards $x\to +\infty$.
The potential $\bar{V}$ needs to be local and short-range. Otherwise, the decay may be non-exponential \cite{winter}.
The energy spectrum is here unbounded from below. In this way, we ignore the problem of finite temperature or bound states that could overlap $|\Omega\rangle$.
Nevertheless, an appropriate $\bar{V}$ can model various physical situations. When $\bar{V}$ is real and antisymmetric, the state does not decay at all.

It gives the closed integral equation for $\Gamma$ from (\ref{gam}),
\begin{eqnarray}
&&\Gamma=\int \bar{V}(y)\bar{V}^\ast(x)e^{\Gamma(x-y)}\theta(x-y)dydx\nonumber\\
&&=\int \frac{dk}{2pi}\frac{|V(k)|^2}{\epsilon-\Gamma-ik}
\end{eqnarray}
for $V(k)=\int dk e^{ikx}\bar{V}(x)$. Here the Fourier formula works in only by the formal expansion
\begin{equation}
(\epsilon-\Gamma-ik)^{-1}=\sum_{n\geq 1}\frac{\Gamma^{n-1}}{(\epsilon-ik)^{n}}.
\end{equation}
In the case of multiple solutions we take that with the smallest $\mathrm{Re}\;\Gamma$.
The divergence of the state $|\psi\rangle=|\Omega\rangle+\int dx\psi(x)|x\rangle$ exhibits in the asymptotic $\psi(x)\sim e^{\Gamma x}$.
The full derivation is given in  Appendix \ref{appa}.

In the lowest order of $V$ we get $R\simeq |V(0)|^2$ (assuming $V(0)\neq 0$), $\langle a(t)\rangle=e^{-Rt}$ and, for $\lambda\to 0$,
\begin{equation}
\langle a(0)||a(t)\rangle_Q\simeq 1+|V(0)|^2t/2-\int dk\frac{|V(k)|^2\sin^2(kt/2)}{\pi k^2} \label{pt0}
\end{equation}
 (see the derivation in  Appendix \ref{appb}), violating (\ref{clas}) for $t> 0$
For long $t$, $\langle a(0)||a(t)\rangle-1$ saturates to
\begin{equation}
\int \frac{dk}{2\pi k^2}(|V(0)|^2-|V(k)|^2),\label{sat}
\end{equation}
taking the Cauchy principal value at $k=0$.

\begin{figure}
\includegraphics[scale=.8]{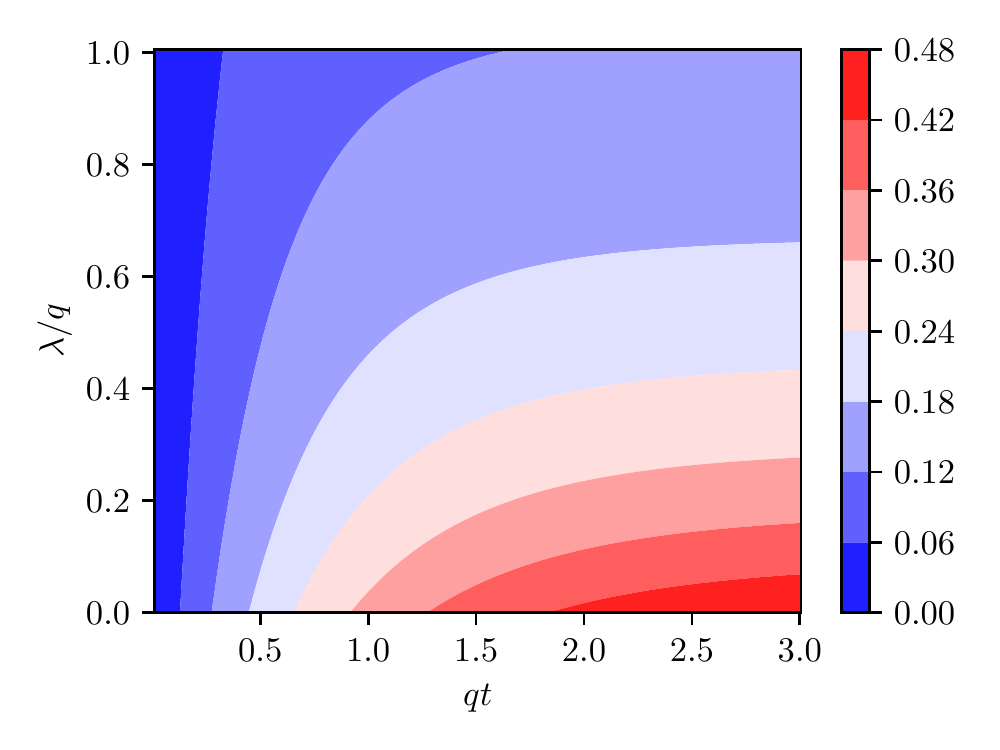}
\includegraphics[scale=.8]{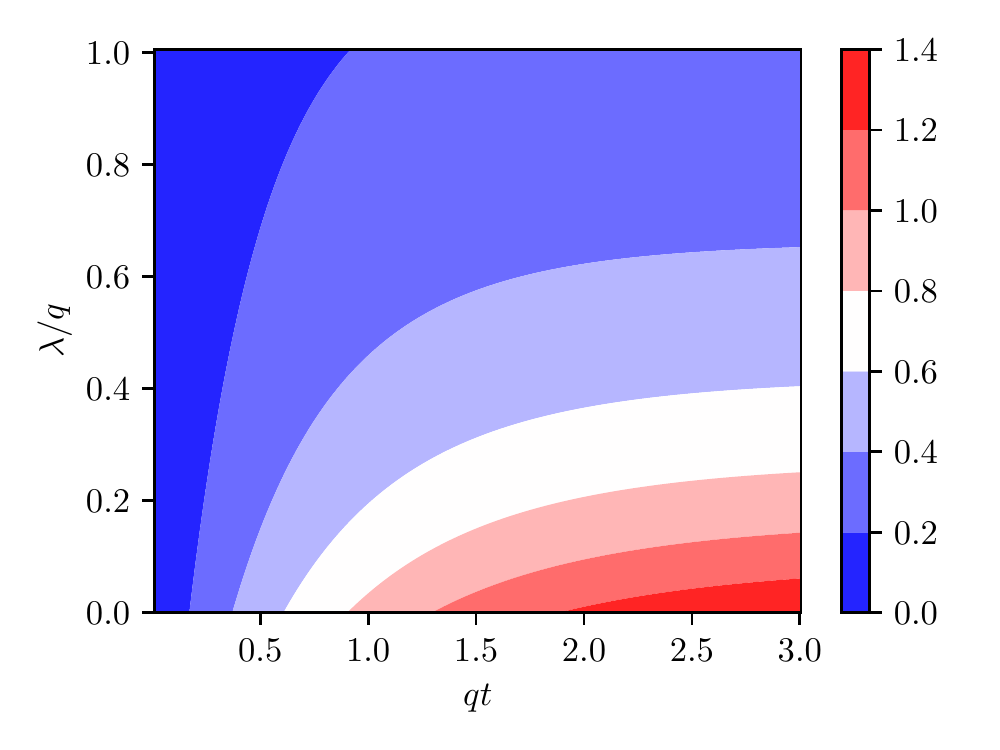}
\caption{The value of $(\langle a(0)||a(t)\rangle_Q-1)/\alpha$ given by (\ref{cor2}) for $|V(k)|^2=\alpha q^{2m-1}/(k^2+q^2)^m$ and $m=1$ (upper), $m=2$ (lower)}
\label{figla}
\end{figure}

\begin{figure}
\includegraphics[scale=.8]{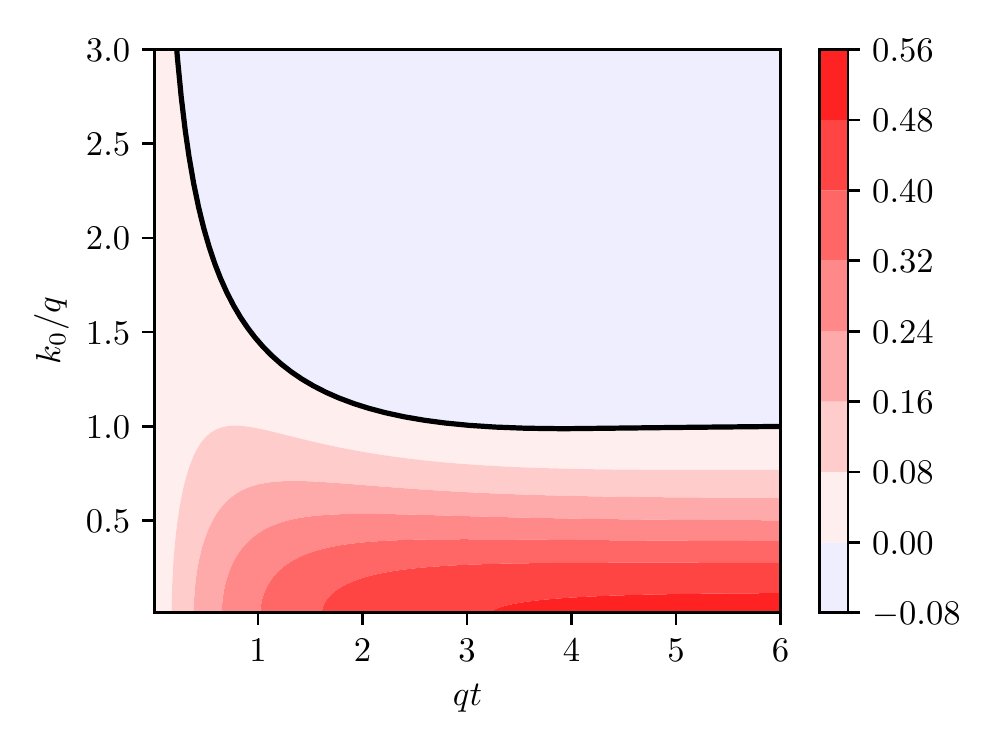}
\includegraphics[scale=.8]{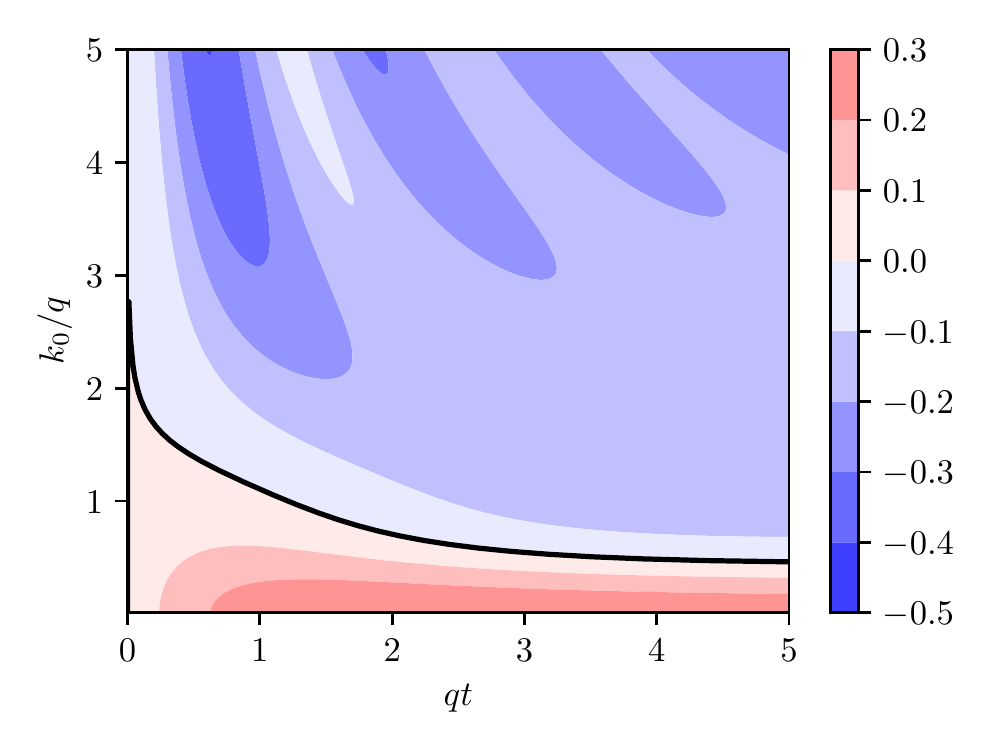}
\caption{The value of $(\langle a(0)||a(t)\rangle-1)/\alpha$ given by (\ref{pt0}) for $|V(k)|^2=\alpha q^{2m-1}/((k-k_0)^2+q^2)^m$, $\lambda=0$ and $m=1$ (upper), $m=2$ (lower).
The black line indicates the value $0$.}
\label{fig0}
\end{figure}

One can also track the violation with  finite $\lambda$, taking into account the modification of the decaying state and its decay rate
(Appendix \ref{appc}),
\begin{eqnarray}
&&R\simeq \int \frac{\lambda|V(k)|^2dk/\pi}{\lambda^2+k^2},\nonumber\\
&&\langle a(0)||a(t)\rangle_Q-1\simeq\label{cor2}\\
&&\mathrm{Re}\int \left(1-e^{-(\lambda-ik)t}\right)\frac{|V(k)|^2dk}{2\pi(\lambda-ik)^2}.\nonumber
\end{eqnarray}
Note that the violation remains for every $\lambda$, even large, at a sufficiently small time. For a long time behavior  one
neglects the second, exponentially decaying term.
For small $\lambda$ the correction to $R$ is just $-2\lambda$ times (\ref{sat}). As the sign of this correction indicates whether the decay is
accelerated or slowed down \cite{kofman,koshino2004}, we see that the violation in the long time limit is equivalent to slowing down the decay, i.e.
quantum Zeno effect. This equivalence is not necessarily true at very strong measurements.
We have plotted a violation of (\ref{cor}) using (\ref{cor2}) for the generic potential 
\begin{equation}
|V(k)|^2=\alpha q^{2m-1}/((k-k_0)^2+q^2)^m\label{lore}
\end{equation}
in the lowest order of $\alpha$ with $m=1,2$.
The case $k_0=0$ and finite $\lambda$ is depicted in Fig.  \ref{figla} 
while $\lambda=0$ and arbitrary $k_0$ is shown in Fig. \ref{fig0}. Note that
the violation occurs only in a particular range of parameters, which can be an effect of a more complicated initial state of the decay (e.g. nonmonotonic).

In the Lorentzian case (\ref{lore})  with $k_0=0$, $m=1$, one can analyze the case exactly, owing to the mapping into pseudomodes \cite{pseudo}
\be
\int dp |V(p)|^2e^{ipt}=\pi \alpha e^{-qt}
\ee
for $t>0$
leading to an equivalent Hamiltonian 
\be
\hat{H}=\begin{pmatrix}
0&\tilde{\alpha}\\
\tilde{\alpha}&-iq\end{pmatrix}
\ee
in the basis $|\Omega\rangle$, $|1\rangle$
where $\tilde{\alpha}^2=\pi\alpha$ and $|1\rangle$ is an auxiliary unphysical state (pseudomode) equivalent to the dynamics of the full reservoir. For the measurement of $\hat{A}$, the corresponding Lindblad operator (\ref{lind}) needs a pseudomode replacement
$[\hat{H},\hat{\rho}]\to
\hat{H}\hat{\rho}-\hat\rho\hat{H}^\dag$. 
The evolution does not affect the antisymmetric part $\hat{\rho}-\hat{\rho}^T$ so we can exclude it from consideration, 
leaving the three independent entries in $\hat{\rho}$. It is convenient to make a shift $\check{L}=\check{L}'-q$. 
In the noninvasive limit $\lambda=0$, the dynamics splits into a left and right evolution (by $\hat{H}$ and $\hat{H}^\dag$)
so $\check{L}'$ has three eigenvalues $0$, $\pm \sqrt{4\pi\alpha-q^2}$. 
For $4\pi\alpha\leq \pi q^2$ the relevant decaying state is the slowest one, 
i.e. for $\check{L}=-q+\sqrt{4\pi\alpha-q^2}$. For $4\pi\alpha>q^2$, two of the eigenvalues become complex, and the system is more as a two-level state than a decaying one.
For a finite $\lambda$ the eigenvalues $L'=\check{L}'$ satisfy
\be
-4\pi\alpha L'-(L^{\prime 2}-q^2)(L'+\lambda)=0,\label{eig}
\ee
which can be factored into
\be
(L'-r)(L^{\prime 2}+(\lambda+r)L'+\lambda q^2/r)\label{eig2}
\ee
with
$4\pi\alpha=(\lambda/r+1)(q^2-r^2)$.
The unique real value of $r\in [0,q]$ always exists for a given $\alpha,\lambda>0$ and can be determined by e.g. numerical solving
of the cubic equation. In all cases the slowest decay is for $L'=r$.
Note that the freezing of the dynamics,
 $r\sim q$, corresponds to $4\alpha/\pi(\lambda+q)\to 0$ which occurs e.g. at $\alpha\to 0$ or $\lambda\to\infty$. The first case
 is due to obvious decoupling for the state $|\Omega\rangle$ but the latter one is the Zeno effect, slowing the decay
 by a sufficiently strong measurement.

The other eigenvalues of $\check{L}'$ read
\be
-(\lambda+r)/2\pm\sqrt{(\lambda+r)^2/4-q^2\lambda/r}.
\ee
Note that they can become complex but the real part is always negative (and so smaller than $r$). The conditional average reads then
\be
\langle a(0)||a(t)\rangle_Q=\frac{\mathrm{Tr}\hat{A}e^{t\check{L}}\check{A}\hat{\rho}}{\mathrm{Tr}\hat{A}\hat{\rho}},\label{cond}
\ee
where $\check{A}=\{\hat{A},\cdot\}/2$, while $\hat{\rho}$ is the eigenstate of $\check{L}$ with $L'=r$. 
The calculation of $e^{t\check{L}}$ can be done using, e.g., the Cayley-Hamilton theorem (Appendix \ref{appd}).
The results, violating the inequality (\ref{clas}), are presented in Fig. \ref{fig2} for $\lambda=0$ and $0.01$, 
using the PYTHON script presented in the Supplemental Material \cite{supp}.

\begin{figure}
\includegraphics[scale=.8]{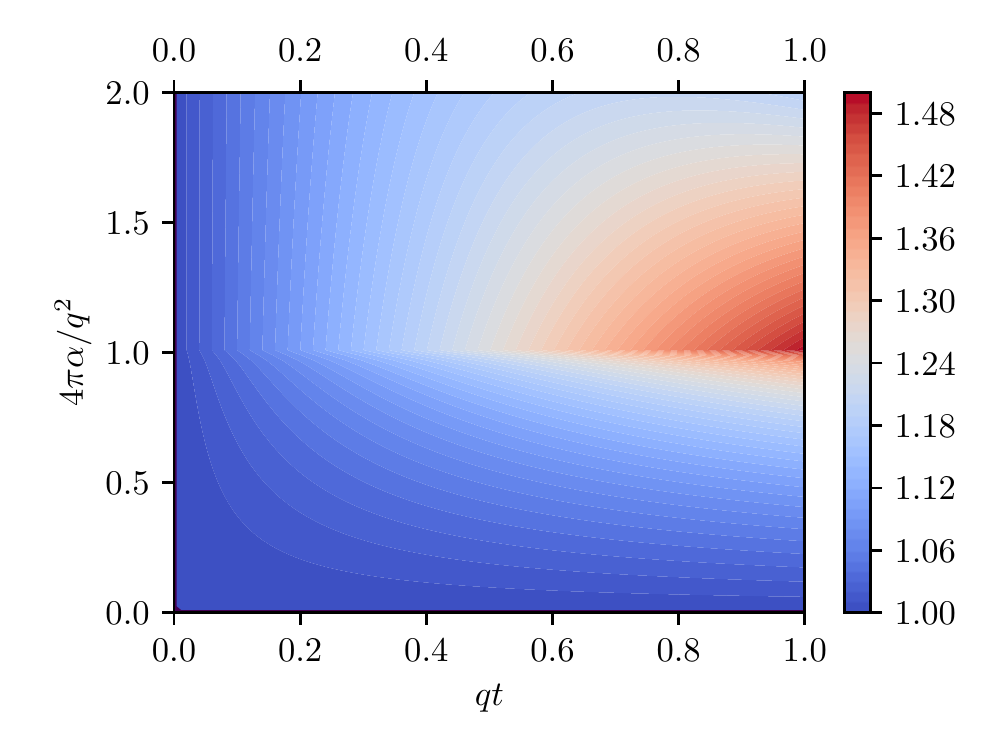}\\
\includegraphics[scale=.8]{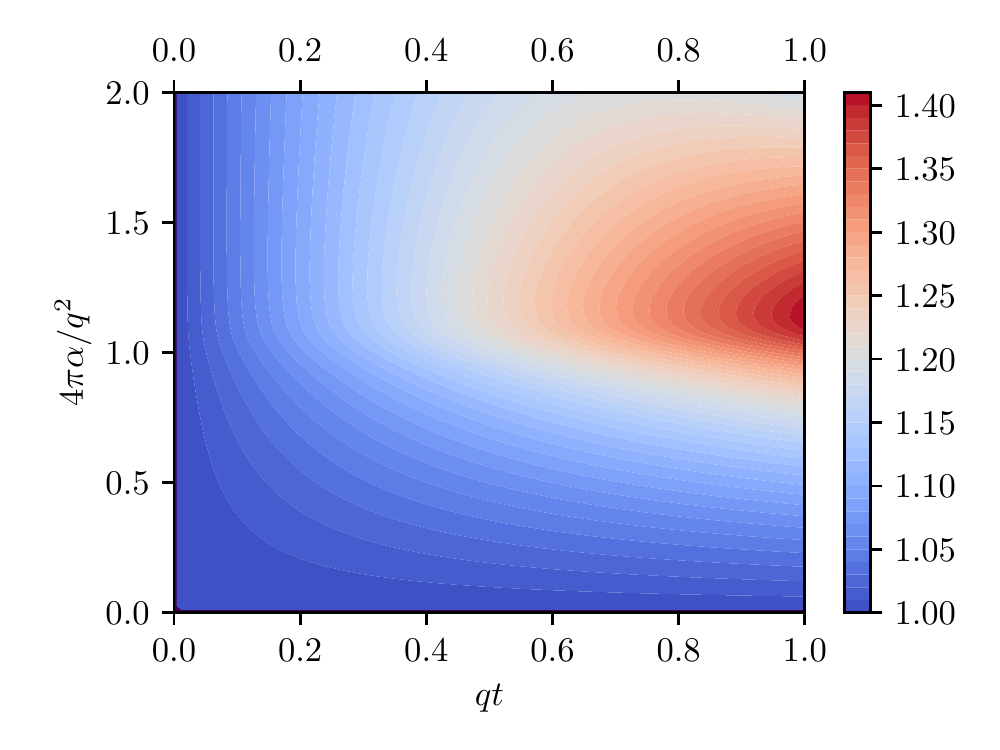}
\caption{The value $\langle a(0)||a(t)\rangle_Q$ for the Lorentzian (\ref{lore}) and the strength of measurement $\lambda=0$ (top) and $0.01$ (bottom),
exceeding the classical limit given by (\ref{clas}) and (\ref{cor}).
Note the nonanalytic behavior at $4\pi\alpha=q^2$ for $\lambda=0$, due to the degeneracy of eigenvalues. }
\label{fig2}
\end{figure}

Apart from the presented example, the violation can be observed in any decaying system, but
the time to make the measurement will be limited by
the non-ideal projection which can partially overlap the ground state. Even in the case of perfect projection, 
the residual occupancy of the decaying state can occur also at finite temperature
or any active reservoir (we assumed a perfect projection and zero temperature).
Nevertheless, most physical decaying systems have well separated timescales so that
the decay is sufficiently slow to allow weak measurements before reaching the vacuum stationary state. 
One can also try the traditional weak value approach \cite{aav}: first, a weak measurement, then projection. The problem can be the high invasiveness of the
projection, which would have to occur shortly after the weak measurement.
The proposal should then be feasible, similarly to already performed experiments on quantum decay \cite{sid1,len,sid2,weakd1}.

\begin{figure}
\includegraphics[scale=0.5]{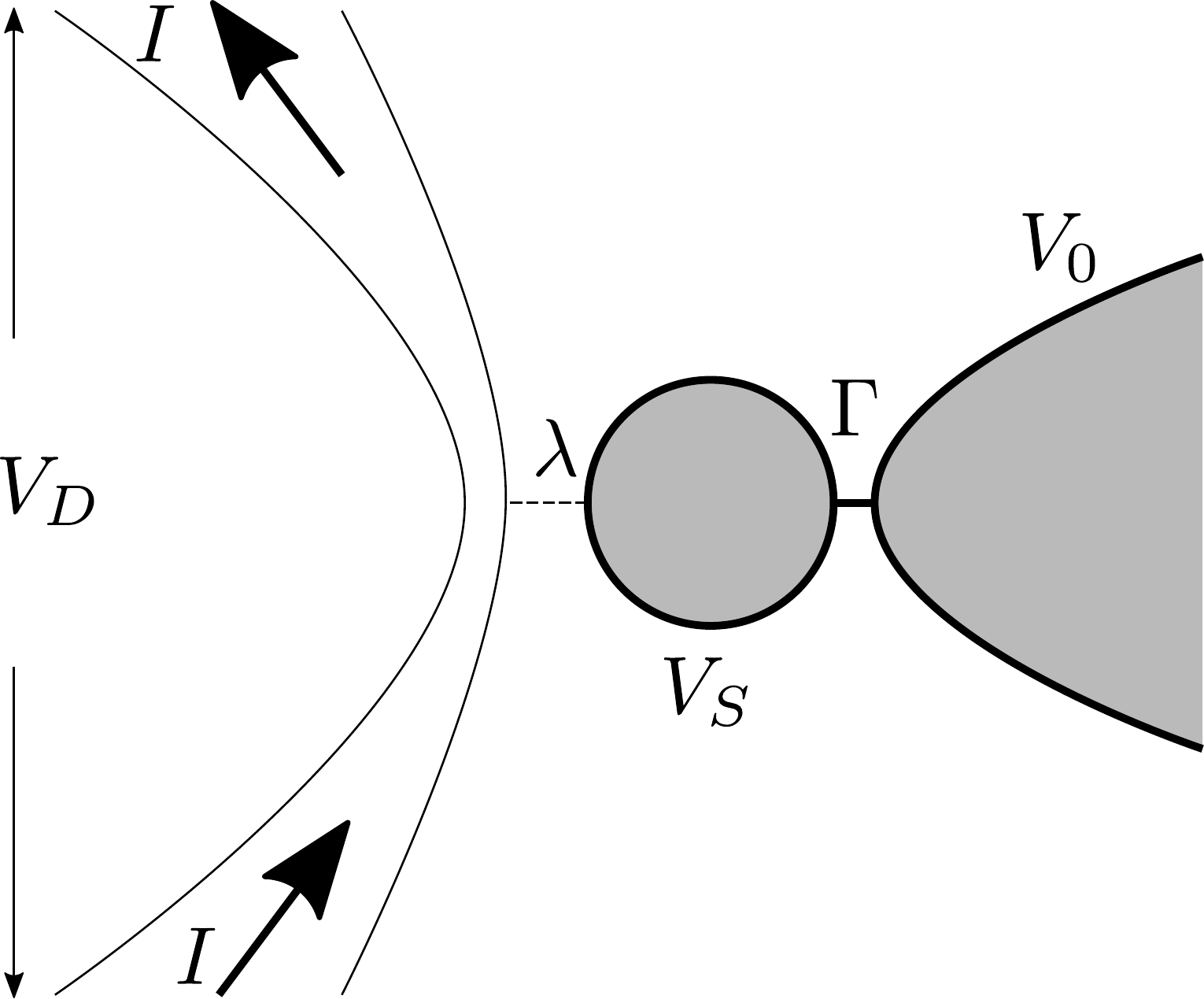}
\caption{A proposed setup to test LG inequalities in a decaying system. The electric current $I$ through a quantum junction at the constant voltage $V_D$
is proportional (by the factor $\lambda$) to the charge occupation $\hat{A}$ of the quantum dot in the middle. The charge in the dot decays to the reservoir 
(decay factor $\Gamma$) on the right. The dot and reservoir are controlled by voltage $V_S$ and $V_0$, respectively.}
\label{dot}
\end{figure}

One of the possible implementations, depicted in Fig. \ref{dot}, could be a quantum dot coupled to a reservoir, which is roughly described by eqs. (\ref{hamp}).
The charge occupation $\hat{A}$ can be measured by a nearby junction whose conductance $I/V_D$ depends linearly on $\lambda\hat{A}$. By monitoring continuously
$I(t)$ one can find correlations $\langle I(t)I(0)\rangle\sim \langle\hat{A}(t)\hat{A}(0)\rangle$. The dot's charge decays into the reservoir.
Controlling the reservoir's voltage $V_0$ once can prepare the dot in the initial, fully occupied state if $V_0\ll V_S$ for the dot's voltage $V_S$, assuming
the negative charge carriers. Switching to $V_0\gg V_S$ the decay starts, allowing the measurement of the correlations. 
Repeating the prepare-and-measure process, the sufficiently large statistics can be gained to 
test the inequalities  (\ref{cor}) and (\ref{acor}).

\section{Coherence time estimation}

The measurement of the correlation $\langle a(0)||a(t)\rangle$ gives useful information about the coherence time of the decay process.
The time to saturate from (\ref{pt0}) to (\ref{sat}) is of the order $1/k_c$ where $k\sim k_c$ is the momentum scale at which $V(k)$ starts to
change (i.e. $V(k\ll k_c)\simeq V(0)$). 
One can even extract the interaction potential
\begin{equation}
|V(k)|^2+|V(-k)|^2=-\int_0^\infty 4\cos(kt)\frac{d^2\langle a(0)||a(t)\rangle}{dt^2}dt
\end{equation}
in the lowest order, directly from (\ref{pt0}).
In the case of environmental decoherence, we can use the Lindblad equation (\ref{lind}) and the results
(\ref{cor2}). Here $\lambda$ has a different role: not as the measurement strength, which is small, but quantifying decoherence. In the case of strong interaction $\lambda\gg k_c$ we get the relation
\begin{equation}
\langle a(0)||a(t)\rangle-1\simeq R(1-e^{-\lambda t})/2\lambda
\end{equation}
which saturates to $R/2\lambda$ at $t\sim 1/\lambda$. Information about the coherence time can help in the practical use of decaying states as a quantum resource,
in combination with the target space of the decay, here restricted in position space to $|x|\lesssim 1/k_c$. Such a combination of the decaying state and
coherent environment forms a kind of partial qubit, with limited coherence and information capacity, but it can be increased by using multiple identical states.
Note that the decoherence time is relative, i.e., the state maintains coherence counted from any starting time point.

\section{Conclusions}

The violation of inequality (\ref{cor}) and (\ref{acor}) is not the only signature of the nonclassicality of decaying states.
One can probe multi-time correlation or generalize it for more complicated states, e.g. with a connected pair of decaying states.
Correlations can be measured by two independent detectors to avoid noise. In principle, the violation requires measuring a projective operator
or its approximation. The violation for other observables may be still possible but needs the detailed analysis of (\ref{aha}).
The fact that the decaying system exhibits nonclassical properties shows its usefulness as a quantum resource. Although our nonclassicality
measure is small for an individual system, 
using a coherent collection of such systems should make it possible to encode quantum information, to be investigated in future research.

\section*{Acknowledgement}

We thank Wolfgang Belzig for useful comments and advice.

\appendix

\section{Example of decaying system}
\label{appa}

Solving Eq.  (\ref{gam}) for $\hat{H}$ given by (\ref{hamp}) for the state $|\psi\rangle=|\Omega\rangle+\int dx\psi(x)|x\rangle$  we denote
$
\psi(x)=e^{\Gamma x}\phi(x)
$
to get
\ba
&&i\partial_x\phi(x)=\bar{V}(x)e^{-\Gamma x},\nonumber\\
&&
\phi(x)=\int_{-\infty}^x \bar{V}(y)e^{-\Gamma y}dy/i,\\
&&
\psi(x)=\int_{-\infty}^x \bar{V}(y)e^{\Gamma(x-y)}dy/i,\nonumber
\ea
which gives a closed equation (\ref{hamp}). Similarly we can find the antidecaying state
$|\tilde\psi\rangle=|\Omega\rangle+\int dx\tilde{\psi}(x)|x\rangle$ as the eigenstate of (\ref{ant}). Analogously we introduce
$
\tilde{\psi}(x)=e^{-\Gamma^\ast x}\tilde\phi(x)
$
to get
\ba
&&i\partial_x\tilde{\phi}(x)=\bar{V}(x)e^{\Gamma^\ast x},\nonumber\\
&&
\tilde{\phi}(x)=-\int_x^\infty \bar{V}(y)e^{\Gamma^\ast y}dy/i,\\
&&
\tilde{\psi}(x)=-\int_x^\infty \bar{V}(y)e^{\Gamma^\ast (y-x)}dy/i.\nonumber
\ea
This gives the same integral equation for $\Gamma$ that we obtained with the decaying state, but $\tilde{\psi}$ diverges in the $x\to -\infty$ direction.

In the momentum representation we have
\be
\Gamma=\int_0^\infty dz\int |V(k)|^2e^{z(\Gamma+ik)}dk/2\pi.
\ee
Changing the integration order is possible only by a formal expansion,
\be
e^{z(\Gamma+ik)}\to e^{-(\epsilon-ik)z}\sum_{n\geq 0}\Gamma^nz^n.
\ee

\section{Continuous measurement -- correlations}
\label{appb}

We are interested in the weak correlation
\be
\langle \{\hat{A}(t),\hat{A}(0)\}\rangle/2=e^{-Rt}g(t),
\ee
taking into account that the state perturbed by $\hat{A}(0)$ will ultimately decay with the same rate as the selfconsistent pseudostationary state.
We start from the auxiliary matrix (not a positive state)
\ba
&&\{\hat{A},\hat{\rho}\}/2=|\Omega\rangle\langle\Omega|+\nonumber\\
&&\int dx\rho(x)|x\rangle\langle\Omega|/2+\int dx\rho^\ast(x)|\Omega\rangle\langle x|/2
\ea
with $\rho(x)$ given by the previous selfconsistent solution.
Now, we apply the Lindblad evolution equation to this matrix to find the projection onto $\Omega$ at time $t$.
Note that
\ba
&&\{\hat{A},\hat{\rho}\}/2=\hat{\rho}-\int dx\rho(x)|x\rangle\langle\Omega|/2\\
&&-\int dx\rho^\ast(x)|\Omega\rangle\langle x|/2
-\int dxdy\rho(x,y)|x\rangle\langle x|\nonumber
\ea
The evolution of $\hat{\rho}$ is by construction just the decay $e^{-Rt}$. On the other hand, the remaining terms are small in 
terms of the perturbative approach. In this way we can find the function $g$ perturbatively. In the lowest order it reads
\ba
&&g(t)=1+ 2\mathrm{Re}\int idx \int_0^t ds \bar{V}^\ast(x)\rho(x-s)e^{-\lambda s}=\nonumber\\
&&1+2\mathrm{Re}\int_0^t ds\int_0^\infty dz \int dx\times\nonumber\\
&&\bar{V}^\ast(x)\bar{V}(x-z-s)e^{-\lambda (z+s)}\nonumber\\
&&=1+\mathrm{Re}\int \frac{|V(k)|^2dk/\pi}{(\lambda-ik)^2}
(1-e^{-(\lambda-ik)t})
\ea

\section{Continuous measurement - decaying rate}
\label{appc}

The Lindblad equation (\ref{lind}) for $\hat{H}$ given by (\ref{hamp}), $\hat{A}=|\Omega\rangle\langle\Omega|$ and the state
$\hat{\rho}(t)=e^{-Rt}\hat{\rho}$
with
\ba
&&\hat{\rho}=|\Omega\rangle\langle\Omega|+\int dx\rho(x)|x\rangle\langle\Omega|+\\
&&\int dx\rho^\ast(x)|\Omega\rangle\langle x|+\int dxdy\rho(x,y)|x\rangle\langle y|\nonumber
\ea
reads
\ba
&&
-iR=\int dx \bar{V}^\ast(x)\rho(x)-\int dx \bar{V}(x)\rho^\ast(x),\nonumber\\
&&-iR\rho(x)=-i\partial_x\rho(x)+\bar{V}(x)\nonumber\\
&&-\int dy\bar{V}(y)\rho(x,y)-i\lambda\rho(x),\\
&&
-iR\partial_t\rho(x,y,t)=-i\partial_x\rho(x,y)-i\partial_y\rho(x,y)\nonumber\\
&&+\bar{V}(x)\rho^\ast(y)
-\bar{V}^\ast(y)\rho(x).\nonumber
\ea
The decay solution should diverge $\rho(x,y)\to g(x-y)e^{R(x+y)/2}$ for some $g$ so
\ba
&&i\rho(x,y)
=\int_0^{\infty} e^{Rz}dz\times\nonumber\\
&&(\bar{V}(x-z)\rho(y-z)-\bar{V}(y-z)\rho(x-z)),\\
&&
i\rho(x)=\int_0^{\infty} \bar{V}(x-z)e^{(R-\lambda)z}\varrho dz\nonumber\\
&&-\int_0^{\infty}e^{(R-\lambda)z}dz \int\rho(x-z,y)\bar{V}(y)dy.\nonumber
\ea
In the lowest order for $V$ we have $\rho^{(0)}(x,y)=\rho^{(0)}(x)=R^{(0)}=0$ and 
\be
i\rho^{(1)}(x)=\int_0^{\infty} \bar{V}(x-z)e^{-\lambda z}dz
\ee
and $\rho^{(1)}(x,y)=R^{(1)}=0$ which gives 
\ba
&&R^{(2)}=\int_0^\infty e^{-\lambda z}dzdx\times\nonumber\\
&&(\bar{V}^\ast(x) \bar{V}(x-z)+\bar{V}(x)\bar{V}^\ast(x-z)) \nonumber\\
&&=\int \frac{\lambda|V(k)|^2dk/\pi}{\lambda^2+k^2}.
\ea

\section{Lorentzian case}
\label{appd}

The density matrix,
\be
\hat{\rho}=\begin{pmatrix}
\rho_{00}&\rho_{01}\\
\rho_{10}&\rho_{11}\end{pmatrix}
\ee
undergoes evolution given by 
\be
\check{L}\begin{pmatrix}
\rho_{00}\\
\rho_{01}\\
\rho_{10}\\
\rho_{11}
\end{pmatrix}=\begin{pmatrix}
0&iA&-iA&0\\
iA&-g-B&0&-iA\\
-iA&0&-g-B&iA\\
0&-iA&iA&-2B
\end{pmatrix}
\begin{pmatrix}
\rho_{00}\\
\rho_{01}\\
\rho_{10}\\
\rho_{11}
\end{pmatrix}
\ee
and the measurement operator reads
\be
\check{\Omega}\begin{pmatrix}
\rho_{00}\\
\rho_{01}\\
\rho_{10}\\
\rho_{11}
\end{pmatrix}=\begin{pmatrix}
\rho_{00}\\
\rho_{01}/2\\
\rho_{10}/2\\
0
\end{pmatrix}
\ee
Note also that the simple effect of the final operation $\mathrm{Tr}\check{\Omega}\hat{\rho}=\rho_{00}$.
One can quickly notice that $\rho_{01}=-\rho_{10}$ all the time if it is satisfied initially.
Since we can assume the undecayed state far in the past, i.e. $\rho_{01}=\rho_{10}=\rho_{11}=0$ for $t\to -\infty$,
the four-dimensional space reduces effectively to three dimensions, denoting $X=\rho_{00}$, $Y=i\sqrt{2}\rho_{01}=-i\sqrt{2}\rho_{10}$, $Z=\rho_{11}$.
so that
\be
\check{L}\begin{pmatrix}
X\\
Y\\
Z
\end{pmatrix}=\begin{pmatrix}
0&\sqrt{2}\tilde{\alpha}&0\\
-\sqrt{2}\tilde{\alpha}&-\lambda-q&\sqrt{2}\tilde{\alpha}\\
0&-\sqrt{2}\tilde{\alpha}&-2q
\end{pmatrix}
\begin{pmatrix}
X\\
Y\\
Z
\end{pmatrix}
\ee
and
\be
\check{A}\begin{pmatrix}
X\\
Y\\
Z
\end{pmatrix}=\begin{pmatrix}
X\\
Y/2\\
0
\end{pmatrix}
\ee

For $\lambda=0$, the evolution can be calculated as
\ba
&&e^{(q+\check{L})t}=\hat{1}+qt\sinc\Delta t
\begin{pmatrix}
q&\sqrt{2}\tilde{\alpha}&0\\
-\sqrt{2}\tilde{\alpha}&0&\sqrt{2}\tilde{\alpha}\\
0&-\sqrt{2}\tilde{\alpha}&-q
\end{pmatrix}+\nonumber\\
&&2q^2t^2\sinc^2\frac{\Delta t}{2}
\begin{pmatrix}
q^2-2\tilde{\alpha}^2&\sqrt{2}\tilde{\alpha}q&2\tilde{\alpha}^2\\
-\sqrt{2}\tilde{\alpha}q&-4\tilde{\alpha}^2&-\sqrt{2}\tilde{\alpha}q\\
2\tilde{\alpha}^2&\sqrt{2}\tilde{\alpha}q&q^2-2\tilde{\alpha}^2
\end{pmatrix}
\ea
for $\Delta=\sqrt{4\pi\alpha^2-q^2}$.
Here $\sinc(z)=\sin z/z$ for all complex  $z\neq 0$ and $\sinc(0)=1$.

In the general case, the eigenstate of $L=r-q$ reads
\be
\begin{pmatrix}
X\\Y\\Z\end{pmatrix}=\begin{pmatrix}
\sqrt{2}\tilde{\alpha}/(q-r)\\
-1\\
\sqrt{2}\tilde{\alpha}/(q+r)
\end{pmatrix} \label{statg}
\ee
%Note that it agrees with (\ref{stat}) at $\lambda\to 0$ if $4\pi\alpha>q^2$. On the other hand
%at $4A^2<1$ and $g\to 0$ we have $r\to 0$ leading to a very large $X$.

By the Cayley-Hamilton theorem, we can write again
\ba
&&e^{\check{L}t}=(a+b(\lambda+r)/2+c(\lambda+r)^2/4)\hat{1}+\nonumber\\
&&(b+c(\lambda+r))\begin{pmatrix}
q&\sqrt{2}\tilde{\alpha}&0\\
-\sqrt{2}\tilde{\alpha}&-\lambda&\sqrt{2}\tilde{\alpha}\\
0&-\sqrt{2}\tilde{\alpha}&-q
\end{pmatrix}+\\
&&c(t)\begin{pmatrix}
q^2-2\tilde{\alpha}^2&\sqrt{2}\tilde{\alpha}(q-\lambda)&2\tilde{\alpha}^2\\
\sqrt{2}\tilde{\alpha}(\lambda-q)&\lambda^2-4\tilde{\alpha}^2&-\sqrt{2}\tilde{\alpha}(\lambda+q)\\
2\tilde{\alpha}^2&\sqrt{2}\tilde{\alpha}(\lambda+q)&q^2-2\tilde{\alpha}^2
\end{pmatrix}\nonumber
\ea
with
\ba
&&b(t)=e^{-t(q+(\lambda+r)/2)}t\sinc\Delta t\nonumber\\
&&c(t)=\frac{e^{-t(q+(\lambda+r)/2)}}{2r^2+r\lambda+\lambda q^2/r}\times\nonumber\\
&&(e^{(3r+\lambda)t/2}-qt\sinc\Delta t(3r+\lambda)/2-\cosh\Delta t),\nonumber\\
&&
a(t)=e^{-t(q+(\lambda+r)/2)}\cosh\Delta t-c(t)\Delta^2,\nonumber\\
&&
\Delta=\sqrt{(\lambda+r)^2/4-\lambda q^2/r},
\ea
including the imaginary $\Delta$.

%\end{widetext}

\end{document}